\documentclass[aoas,preprint]{imsart}

\RequirePackage{amsthm,amsmath,amsfonts,amssymb}
\RequirePackage[authoryear]{natbib}
\RequirePackage[colorlinks,citecolor=blue,urlcolor=blue]{hyperref}
\RequirePackage{graphicx}

\startlocaldefs

\theoremstyle{remark}

\endlocaldefs

\begin{document}

\begin{frontmatter}
\title{A Finite Mixture Hidden Markov Model for Intermittently Observed Disease Process with Heterogeneity and Partially Known Disease Type}
\runtitle{MHMM for Disease Heterogeneity with Partially Known Disease Type}

\begin{aug}

\author[A]{\fnms{Yidan} \snm{Shi}\ead[label=e1,mark]{yidan.shi@nyulangone.org}\orcid{0000-0002-3190-6724}},
\author[B]{\fnms{Leilei} \snm{Zeng}\ead[label=e2,mark]{lzeng@uwaterloo.ca}\orcid{0000-0002-2336-7249}},
\author[B]{\fnms{Mary E.} \snm{Thompson}\ead[label=e3,mark]{methompson@uwaterloo.ca}\orcid{0000-0003-2354-8880}}
\and
\author[C]{\fnms{Suzanne L.} \snm{Tyas}\ead[label=e4,mark]{suzanne.tyas@uwaterloo.ca}\orcid{0000-0003-3007-3940}}

\address[A]{Division of Biostatistics, Department of Population Health, New York University Grossman School of Medicine,
\printead{e1}}
\address[B]{Department of Statistics and Actuarial Science,
University of Waterloo,
\printead{e2,e3}}

\address[C]{School of Public Health Sciences,
University of Waterloo,
\printead{e4}}
\end{aug}

\begin{abstract}

Continuous-time multistate models are widely used for analyzing interval-censored data on disease progression over time. 
Sometimes, diseases manifest differently and what appears to be a coherent collection of symptoms is the expression of multiple distinct disease subtypes. 
To address this complexity, we propose a mixture hidden Markov model, where the observation process encompasses states representing common symptomatic stages across these diseases, and each underlying process corresponds to a distinct disease subtype. 
Our method models both the overall and the type-specific disease incidence/prevalence accounting for sampling conditions and exactly observed death times. 
Additionally, it can utilize partially available disease-type information, which offers insights into the pathway through specific hidden states in the disease process, to aid in the estimation. 
We present both a frequentist and a Bayesian way to obtain the estimates. 
The finite sample performance is evaluated through simulation studies. 
We demonstrate our method using the Nun Study and model the development and progression of dementia, encompassing both Alzheimer's disease (AD) and non-AD dementia.
\end{abstract}

\begin{keyword}
\kwd{Auxiliary information}
\kwd{Disease heterogeneity}
\kwd{Hidden Markov model}
\kwd{Interval censoring}
\kwd{Mixture model}
\kwd{Multistate model}
\end{keyword}

\end{frontmatter}

\section{Introduction}
\label{sec:intro}

Modeling disease process data is one of the most important applications of statistics. 
Traditionally, it is assumed that individuals' symptoms stem from a single disease cause. 
However, what may seem like a unified set of symptoms can represent various distinct disease subtypes, each progressing at its own pace.
Accounting for such disease heterogeneity is essential, while determining the disease subtype can be challenging.
Fortunately, auxiliary data sometimes offer insights for disease typing. 
For instance, genome data may help identify cancer subtypes, while brain imaging or neuropathological assessments aid in diagnosing neurodegenerative diseases. 
In this paper, we propose a solution to such a situation with an application of modeling the disease process of dementia using the Nun Study data.

An overview of the Nun Study \citep{snowdon2003} is provided in Section~\ref{sec:motivation}. 
In brief, the study focused on investigating the onset and progression of dementia. 
Only living individuals were eligible for study participation at baseline. 
Cognitive function assessments were conducted at follow-up visits to ascertain the onset of dementia among participants: the time to disease is therefore interval-censored. 
Many participants died during the follow-up period, and if their cognitive status had been normal at the last visit, their disease status was unknown prior to death.
Standard survival analysis using competing risk models considers the time of onset of the disease to be right-censored at that visit, which, however, may induce an underestimation of disease incidence \citep{leffondre2013interval}. 
Alternatively, we consider multistate model approaches, which appropriately account for the possibility that dementia may or may not occur between the last visit and death. 

Intensity-based multistate models provide a powerful and natural framework for characterizing disease processes and dealing with intermittent observation, loss to follow-up, and other complications arising in complex life history processes \citep{cook2018multistate}. 
Time-homogeneous Markov models stand out as the most widely used as they are mathematically tractable, which simplifies the likelihood evaluation. 
However, such models impose two overly strict distribution assumptions: (1) the sojourn time follows an exponential distribution, and (2) the intensity of transitioning out of a state remains constant over time.
There is a growing focus on scenarios where these assumptions are not met.
Methods employing piecewise constant transition intensities offer a feasible extension to the time-homogeneous Markov model, as the existing methodology can be applied within each time interval \citep{gentleman1994, andersen2002}. 
Semi-Markov models enable transition intensities to vary based on the sojourn time in each state. 
For instance, \cite{satten1999} proposed a semi-Markov model with unknown initiation times, \cite{foucher2005} applied the semi-Markov model with generalized Weibull intensities to the HIV disease process, and 
\cite{kang2006} introduced a semi-Markov model tailored for panel data analysis.
Hidden Markov models (HMM) assume that the observation process arises from a latent, or "hidden," Markov process.
\cite{titman2010} introduced a unique HMM tailored for intermittently observed multistate processes. This approach aims to approximate a semi-Markov process by delineating a phase-type model for the underlying process, incorporating additional hidden states.
However, it does not deal with the complication when the disease process is under a mixed observation scheme: the time of progression is subject to interval censoring while the time of death is subject to right censoring. 
Moreover, it does not account for the response-dependent sampling such that only living individuals are recruited.

The methods discussed above primarily focus on modeling the dynamics of individual disease processes without accounting for heterogeneity. 
However, dementia is a clinical syndrome with various potential causes, including Alzheimer's disease (AD), vascular dementia (VaD), Lewy body dementia (LBD), and others \citep{garcia2014mortality}.
Among these, AD stands out as the predominant subtype, with a notably higher incidence among the older population. 
Similar scenarios are observed in other disease domains, such as diabetes, breast cancer, autoimmune diseases, and neurodegenerative disorders like Parkinson's disease.
In these cases, each subtype may exhibit distinct rates of reaching clinically relevant disease stages, such as symptom onset, progression, and mortality. 
Therefore, it becomes essential to consider this heterogeneity when modeling disease processes and their progression.

There has been a growing interest in addressing disease heterogeneity through the application of mixture hidden Markov models (MHMM).
\cite{altman2007mixed} introduced an MHMM approach using the framework of generalized linear mixed effects models in the longitudinal data setting.
\cite{naranjo2022mixed} discussed a model for binary outcomes with misspecification.
\cite{ghassempour2014clustering} proposed an MHMM to capture health trajectories, utilizing time series as the observation process.
Furthermore, discussions have emerged regarding MHMMs in semiparametric regression models \citep{kang2019bayesian}, longitudinal data analysis  \citep{maruotti2012mixed}, and integration with machine learning techniques \citep{wijeratne2020learning}.
However, a notable gap exists in the literature concerning MHMMs when the observation process involves multistate models with panel observations.

Given that AD is distinguished by the presence of plaques formed by amyloid $\beta$ peptide accumulation and intracellular neurofibrillary tangles composed of hyperphosphorylated $\tau$-protein \citep{o2011amyloid}, differentiating AD from other forms of dementia based on neuropathological assessments proves more reliable than relying solely on clinical cognitive symptoms. 
However, collecting neuropathological samples is costly and invasive.
In the Nun Study, neuropathological assessments of AD pathology were performed on a subset of participants who passed away during the follow-up period \citep{riley2002alzheimer}. 
This subsample provided partial information on disease subtypes. 
To the best of our knowledge, no previously existing method has leveraged such auxiliary information, particularly in settings where they are available only for a subset of participants.

In this article, we introduce an MHMM in continuous time designed to analyze multistate disease processes while accommodating a mixed-type observation scheme and sampling conditions. 
Our methodology employs a finite mixture of phase-type Markov chains to capture the disease heterogeneity using partially available disease-type information to aid the estimation procedure, enhance efficiency, and address identifiability and estimability challenges.
The rest of this paper is organized as follows:
Section~\ref{sec:motivation} provides an overview of the Nun Study and outlines the MHMM framework tailored to estimate the dementia disease process.
Section~\ref{sec:method} presents the general methodology for MHMMs and its adaptation to the Nun Study context.
Section~\ref{sec:sim} details simulation studies conducted to assess the performance of our proposed methods and addresses identifiability and estimability issues in MHMMs with potential solutions.
In Section~\ref{sec:resultsNun}, we apply our proposed models to analyze the Nun Study data, utilizing both frequentist and Bayesian approaches. 
We estimate overall and type-specific prevalence, i.e., the prevalence of AD and non-AD dementia, and compare them to population prevalences reported in  \cite{goodman2017prevalence}.
Finally, Section~\ref{sec:remarks} provides a discussion and directions for future research.
Extensive derivations, simulations and analyses results are presented in Appendices.

\section{Motivation: The Nun Study}
\label{sec:motivation}

The Nun Study \citep{snowdon2003} is a longitudinal investigation of aging and cognition within the School Sisters of Notre Dame, a religious congregation in the United States.
The primary focus was on modeling the onset and progression of dementia. 
At baseline (1991--1993), 678 Catholic sisters, who were at least 75 years old, agreed to participate and consented to brain donation postmortem \citep{riley2002alzheimer}. 
In our analyses, we treated age 75 as the time origin.
Cognitive function assessments were conducted at baseline and approximately annually thereafter for up to 12 assessments or death, whichever occurred first. 
The average follow-up duration was 8.59 years. 
In the event of a participant's death during the study, the age at death was documented. 
Diagnosis of dementia was based on criteria involving memory impairment and deficits in at least one other cognitive domain \citep{riley2002alzheimer, iraniparast2022cognitive}. 
Table \ref{tb:1} provides details on the number of participants diagnosed with dementia and/or who died during the study period.
\begin{table}[b!]
	\caption{Number of dementia and death cases in the Nun Study sample}
	\label{tb:1}
	\centering \begin{tabular}{lccc} 
		\hline \\[-1em]
		& Death & Right-censored & Total \\ \\[-1em]    \hline \\[-1em]
		Dementia & 285 &  12 & 297\\  \\[-1em]
		Dementia-free & 321 &  60 & 381\\   \\[-1em]  \hline \\[-1em]
		Total & 606 &  72 & 678\\   \\[-1em]  \hline 
	\end{tabular}
	
	\noindent 
\end{table}

\begin{figure}[h!]
	\centerline{\includegraphics[width=80ex]{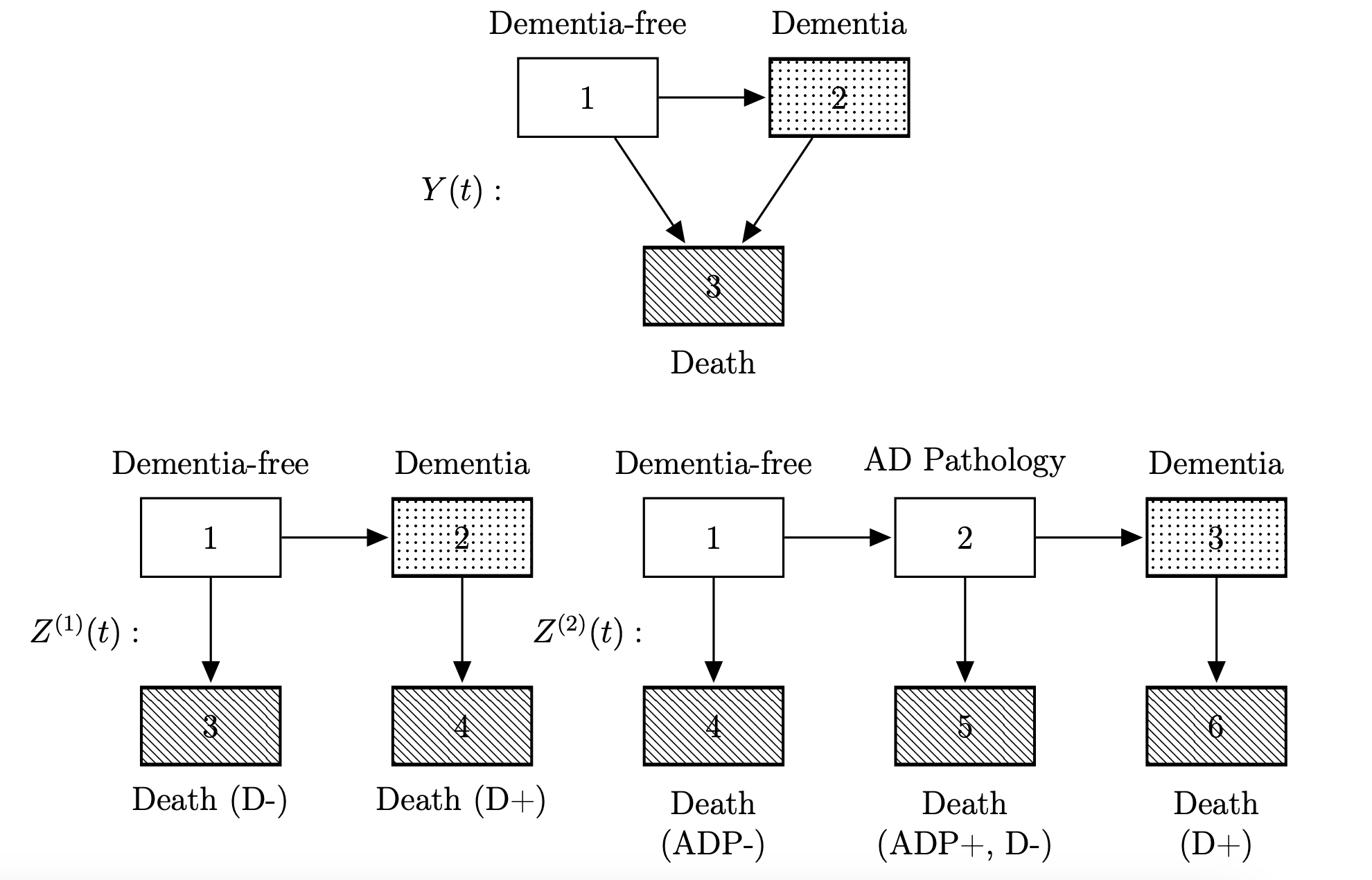}}
	\caption{An MHMM for dementia and Alzheimer's disease. 
		The observation process $Y(t)$ is a three-state process; the underlying process is a mixture of two types of dementia: $Z^{(1)}(t)$ without and $Z^{(2)}(t)$ with the coexistence of AD pathology. 
  ADP+ and ADP- represent with and without AD pathology respectively. 
  D+ and D- represent with and without dementia respectively.
  The filling pattern illustrates the correspondence between underlying states and observed states.}
	\label{fig:model} 
\end{figure}

The dementia process within the Nun Study is conceptualized through a multistate model, as illustrated in Figure~\ref{fig:model}. 
Specifically, the observed disease process based on the clinical assessments, denoted by $Y(t)$,  is characterized by an illness-death model with three states: dementia-free (State 1), dementia (State 2), and death (State 3).
Given that AD constitutes the most prevalent form of dementia, accounting for 60\% to 75\% of cases \citep{ebly1994prevalence, kalaria2008alzheimer, forette1991hypertension}, we assume that the study population is at risk of two distinct types of disease: (1) non-AD dementia and (2) AD. 
Each type is characterized by a unique rate of disease progression. 
Consequently, the underlying disease process is modeled as a mixture of Type I disease (non-AD dementias) denoted by $Z^{(1)}(t)$ and Type II disease (AD) denoted by $Z^{(2)}(t)$.
For $Z^{(1)}(t)$, the underlying disease process is represented by a four-state Markov model. 
For AD ($Z^{(2)}(t)$), we introduce a state representing the onset of AD pathology, which is latent and may not be distinguished in the observed data. 
The pattern coding in Figure \ref{fig:model} illustrates the correspondence between the states in the underlying model and those in the observation model. 
Both processes terminate in absorbing states associated with death in $Y(t)$.
This model assumes that AD pathology precedes the onset of AD-type dementia and that each individual will eventually develop dementia (either AD or non-AD) unless death intervenes before dementia onset.

Diagnosis of AD pathology typically requires substantial resources and access to specialized services such as neuroimaging. 
Therefore, it is often only observed in symptomatic individuals with cognitive impairment who undergo diagnostic testing.
While it is not directly observable in the Nun Study, information on an individual's disease type is partially available. 
Assessment of AD pathology \citep{o2011amyloid} began during the surveillance period and extended beyond the end of the follow-up period. 
As postmortem assessments were still ongoing, our analysis includes neuropathological results for only a subsample of participants (389 participants). 
We have assumed that this subsample is randomly selected, but we will discuss the implications of possible violation of this assumption in Section~\ref{sec:remarks}. 
Among the 389 individuals, 199 were diagnosed with dementia before death. 
Table~\ref{tb:2} presents the number of cases stratified by the presence of AD based on neuropathological assessments and their dementia status determined through clinical evaluation.
Of these cases, 157 were confirmed as AD based on neuropathological evaluation, while 42 were deemed to have non-AD dementia. 
Notably, among the 190 participants who displayed no clinical symptoms of dementia, 106 were found to have AD pathology, suggesting cognitive reserve.
 
\begin{table}[h!]
	\caption{Autopsy results for the pathological evaluation of Alzheimer’s disease (AD) for a subsample of 389 participants in the Nun Study}
	\label{tb:2}
	\centering \begin{tabular}{lccc} 
		\hline \\[-1em]
		& \multicolumn{2}{c}{AD Pathology}  \\ \\[-1em] 
		\cline{2-3} \\[-1em]		
		\multicolumn{1}{c}{Clinical Evaluation} &  Presence  & Absence  & Total \\  \\[-1em] \hline \\[-1em]
		Dementia & 157 & 42 & 199\\  \\[-1em]
		Dementia-free & 106 & 84 & 190\\ \\[-1em]    
		\\[-1em]\hline 
	\end{tabular}
\end{table}

Given the context of the Nun Study, our objective is to model the dementia disease process using the MHMM model in Figure~\ref{fig:model}.
Additionally, we aim to incorporate the partially known neuropathology data presented in Table \ref{tb:2} to enhance the accuracy and efficiency of estimation. 
By the proposed approach, we seek to improve our understanding of the progression and onset of dementia taking into account the distinct characteristics of AD and non-AD dementia subtypes observed in the study population.

\section{Method}
\label{sec:method}

\subsection{Notations}
\label{sec:method_notations}

To simplify notation, we will omit subscripts for individuals. 
Participants were recruited at time $t_{1} \geq 0$, and their disease status was clinically assessed at scheduled times $t_{1}<t_{2}<\ldots<t_{K^{*}}=\tau$, where $\tau$ denotes the administrative censoring time and let $t_0=0$.
Consistent with Section \ref{sec:motivation}, we denote the observed multistate process as ${Y(t); t \ge 0}$. 
At each time point $t_k$, the observed state occupancy of an individual, $Y(t_k)$, is classified as one member of a set of observed statuses $\mathcal{J}_Y={1,2,\ldots,J_Y}$. 
This set may include a subset of absorbing states such as death under various circumstances, represented by $\mathcal{D}_Y$.
The observation process may conclude due to either the participant's dropout at a random time $C$ or their death (or transition to another absorbing state) at a random time $T_{D}$. 
Denote the termination time as $T^{\dagger}=\min\left\{ \tau,C,T_{D}\right\}$ and the absorbing indicator as $\delta=I(Y(T^{\dagger}) \in \mathcal{D}_Y)$. 
The actual number of assessments for a participant is then denoted by $K=\max\left\{ k; t_{k}\leq T^{\dagger},k=0,\ldots,K^{*}\right\} $, where $K= K^{*}$ if the participant was alive and followed until the end of the study, and $K<K^*$ otherwise.
In the scenario mimicking the Nun Study, where the death time is precisely observed, the observational data for a participant who was alive at the end of their follow-up time $T^{\dagger}$ are $\left\{ Y(t_{1}),\ldots,Y(t_{K}),\delta=0\right\} $. 
For a deceased participant, the observational data are $\{Y(t_{1}), \ldots, $ $ Y(t_{K}), T_{D},\delta=1\}$.

In a typical HMM, the state occupancy of the observation process, $\{Y(t); t \ge 0\}$, depends on a latent process $\{Z(t); t \geq 0\}$ that is assumed to be Markovian with a state space $\mathcal{J} = \{1, 2, \ldots, J\}$. 
The latent process $\{Z(t); t \geq 0\}$ can be characterized by initial probabilities and a transition intensity matrix.
The initial probabilities are represented by a row vector $\boldsymbol{\pi} = (\pi_1, \ldots, \pi_J)$, where $\pi_j = P(Z(0)=j )$, $j=1, \ldots, J$.
The transition intensity matrix $\boldsymbol{\Lambda}(t)=   [\lambda_{ij}(t)]$ governs the transitions between states over time.
The ($i, j$)-th off-diagonal element
\begin{equation*}
	\lambda_{ij}(t)=\lim_{\Delta t\rightarrow0}\frac{P(Z(t^-+\Delta t)=j|Z(t^-)=i, \mathcal{H}_Z(t^-))}{\Delta t}, ~i \ne j, ~\mbox{and}~ i, j \in \mathcal{J}
\end{equation*}
represents the instantaneous risk of making a state $i$ to state $j$ transition at time $t$, and the ($i, i$)-th diagonal entry is $-\sum_{j: j\neq i}\lambda_{ij}(t)$.
Due to the Markov assumption, the transition intensity $\lambda_{ij}(t)$ is independent of the history of the process up to the current time point, denoted by $\mathcal{H}_Z(t)=\left\{Z(s);0\leq s< t \right\}$.
Markov models are attractive because transition probabilities $p_{ij}(s,t)=P\left(Z(t)=j|Z(s)=i\right)$, for $i, j \in \mathcal{J}$, can be derived by solving the Kolmogorov forward differential equation. 
Specifically, if we denote the transition probability matrix as $\mathbb{P}(s,t) = [p_{ij}(s,t) ]$, then we have $\partial \mathbb{P}(s, t)/ \partial t = \mathbb{P}(s, t) \boldsymbol{\Lambda}(t)$ for $t > s$.
When the process is time-homogeneous, i.e., $\boldsymbol{\Lambda}(t)=\boldsymbol{\Lambda}$, an explicit solution for the transition probability matrix exists such that $\mathbb{P}\left(s,t\right) = \exp\{(t-s)\boldsymbol{\Lambda}\}$.

A key assumption in HMMs is that at any given time point $t$, the observed state $Y(t)$ is solely determined by the underlying state $Z(t)$ and not by the history of the observed and underlying processes, that is,   
$
P\left(Y(t)=j \mid Z(t)=i,\text{\ensuremath{\mathcal{H}}}_{Y}\left(t^{-}\right),\text{\ensuremath{\mathcal{H}}}_{Z}\left(t^{-}\right)\right) = P\big (Y(t)=j \mid Z(t)=i\big) =  e_{ij},\ i\in\mathcal{J},\ j\in\mathcal{J}_Y, 
$
where $e_{ij}$ is the so-called emission probability which captures the mapping between the observation state and the underlying disease process.
$e_{ij}=1$ if a latent state $i$ is classified as the observed state $j$, and 0 otherwise. 
When there is a one-to-one mapping between the observed and underlying states without any misclassification, the HMM reduces to a regular Markov model.
Let $E= [ e_{ij} ]$ denote the emission probability matrix in general. In the proposed model, the emission probability matrix is pre-defined and reflects a quantified description of the correspondence between the observation model and the underlying model.

\subsection{MHMM for disease processes with heterogeneity}
\label{sec:method_MHMM}

Now suppose participants are susceptible to multiple diseases or disease subtypes, characterized by stochastic processes $Z^{(1)}(t), Z^{(2)}(t),\ldots,$ $Z^{(M)}(t)$ respectively, with $Z^{(m)}(t)$ having state space $\mathcal{J}^{(m)}$, $m=1, 2, \ldots, M$. 
Assume these diseases cannot be distinguished by the stages of clinical symptoms, meaning they share the same observed disease status $Y(t)$. 
Let $\Delta$ represent the disease type. 
The underlying model then takes the form of one of the underlying processes, as follows:
\begin{eqnarray*}
	Z(t) & = & \sum_{m=1}^M \mathbb{I}(\Delta=m)Z^{(m)}(t),
\end{eqnarray*}
where each $\{ Z^{(m)}(t); t \geq 0\}$ is a distinctive Markov process with initial probabilities $\boldsymbol{\pi}^{(m)}$, transition intensity matrix $\Lambda^{(m)}(t)$, and emission probabilities $\{\mathrm{e}_{ij}^{(m)}, i\in\mathcal{J}^{(m)}, j\in\mathcal{J}_Y\}$. 

For the Nun Study case, the dementia-like symptoms can be caused by either AD (the most common cause) or non-AD dementia. 
So, we can think of the study population being at risk of two ``types" of disease: type I for non-AD dementias and type II for AD. 
From this population we obtain clinical assessments of cognitive function, including dementia, approximately annually. 
As shown in Figure \ref{fig:model}, the observed process $Y(t)$ is a three-state process characterizing onset of dementia-like symptoms and occurrence of death with or without the condition; the underlying process is a mixture of type I disease (non-AD dementias) denoted by $Z^{(1)}(t)$ and type II disease (AD) denoted by $Z^{(2)}(t)$. 
For the AD process we introduce a state representing the onset of AD pathology and assume dementia will be caused by this.  
The diagnosis of AD requires substantial resources and access to services such as neuroimaging; in addition, only symptomatic individuals with cognitive impairment typically come to clinical attention and are referred for such diagnostic testing.
So, the state of AD pathology is viewed as a latent state that may not be observable. 
The shades in Figure \ref{fig:model} represents the correspondence of the state(s) in the underlying model to the states in the observation model. 
For $Z^{(1)}(t)$, this underlying disease process is a 4-state Markov process. 
For $Z^{(2)}(t)$, the disease process is a 6-state Markov model where the dementia-free state and the pathology state cannot be differentiated in the observed data. 
The absorbing states for both processes are related to the death state of $Y(t)$.
The emission matrices that connect the observed states and the states of each underlying disease process are
\begin{eqnarray*}
\mathrm{E}^{(1)} = \left[\begin{array}{ccc}
1 & 0 & 0\\
0 & 1 & 0\\
0 & 0 & 1\\
0 & 0 & 1
\end{array}\right]~~\mbox{and}~~~	
\mathrm{E}^{(2)} =  \left[\begin{array}{ccc}
1 & 0 & 0\\
1 & 0 & 0\\
0 & 1 & 0\\
0 & 0 & 1\\
0 & 0 & 1\\
0 & 0 & 1
\end{array}\right]\;.
\end{eqnarray*}
Such a model assumes that AD pathology would precede Alzheimer-type dementia, and that each individual would eventually develop dementia (AD or non-AD dementia) unless death happened before the development of dementia.

The idea for constructing the likelihood is to sum over all possible underlying paths associated with the observed process. 
Let $\psi_{m}=P(\Delta=m)$ represent the mixture probability and $\sum_{m=1}^M \psi_m =1$.
Denote the unknown parameter as $\boldsymbol{\Theta}= \{\psi_m, \boldsymbol{\pi}^{(m)}, \boldsymbol{\Lambda}^{(m)}, m=1,\ldots,M\}$, where $\boldsymbol{\Lambda}^{(m)}$ is the transition intensity parameters for underlying process $m$.
In the simple case without any sampling condition, and with a strictly intermittent observation scheme, the likelihood for the panel observations $\{Y(t_1)=y_1, \ldots, Y(t_K)=y_k\}$ under the MHMM can be expressed in a matrix-multiplication form as follows:
\begin{eqnarray}
\label{eq:Lh0}
\mathcal{L} (\boldsymbol{\Theta})=\sum_{m=1}^M \psi_{m} \boldsymbol{\pi}^{(m)} \mathbb{Q}_{1}^{(m)} \mathbb{Q}_{2}^{(m)}\ldots \mathbb{Q}_{K}^{(m)}\mathbf{1},
\end{eqnarray} 
where
$\mathbf{1}$ represents a column vector of $1$'s with a length equal to the number of states in the state space $\mathcal{J}^{(m)}$
and
\begin{eqnarray*}
\mathbb{Q}^{(m)}_k &=&\mathbb{P}^{(m)}(t_{k-1}, t_{k}) \mathbb{E}^{(m)}(t_k)
\end{eqnarray*}
is a product of the transition probability matrix and a diagonal matrix of emission probabilities $\mathbb{E}^{(m)}(t_k)= \textrm{diag}\{e^{(m)}_{1y_k}, \ldots, e^{(m)}_{J y_k}\}$.

Modifications to the likelihood (\ref{eq:Lh0}) are necessary for disease processes that involve more complex sampling and observation schemes and terminating events such as death. 
For instance, in the Nun Study, individuals who reached the absorbing state (death) before the study's onset were not included in the sample. 
In this case, the sampling condition can be expressed as $Y(t_1)\notin \mathcal{D}_Y$.
For a given disease type $m$ and entry time $t_1$, the probability of being sampled is:
\begin{eqnarray*}
	g(\boldsymbol{\Theta}^{(m)}; t_1) =P\big (Y(t_1) \notin \mathcal{D}_Y \mid M=m \big ) &=& \boldsymbol{\pi}^{(m)}\mathbb{P}^{(m)}\left(0,t_{1}\right)\mathbf{d}^{(m)}
\end{eqnarray*} 
where $\boldsymbol{\Theta}^{(m)}=\{\psi_m, \boldsymbol{\pi}^{(m)}, \boldsymbol{\Lambda}^{(m)}\}$ represents the parameter set for disease type $m$, and $\mathbf{d}^{(m)}$ is a column vector with elements $d^{(m)}_i = I  ( \exists j: j\notin \mathcal{D}_Y, e^{(m)}_{ij}=1  )$ indicating that state $i$ of $Z^{(m)}(t)$ is associated with a non-absorbing state of the observation process $Y(t)$.
The Nun Study also has a mixed observation scheme. 
While dementia status is intermittently assessed, time to death is either exactly observed or right-censored. 
When the age at death ($T_D$) is exactly recorded, the participant can be in any of the non-absorbing states at the immediate moment prior to $T_D$.
Hence, for a given disease type $m$, we define the matrix 
\begin{eqnarray*}
\mathbb{Q}^{(m)}_{D}& = &\mathbb{P}^{(m)} \big ( t_{K},  t_{D} \big) \boldsymbol{\Lambda}^{(m)}(t_D) \mathbb{E}^{(m)}(t_D),
\end{eqnarray*}
where the $(i, j)$-th entry is $e^{(m)}_{j y_{t_D}} \sum_{u=1}^J p^{(m)}_{iu}(t_K, t_D)\lambda^{(m)}_{ij}(t_D)$.
Therefore, the likelihood accounting for the sampling condition and a mixed type of observations is:
\begin{eqnarray}
\label{eq:Lh}
\mathcal{L}(\boldsymbol{\Theta}) =\sum_{m=1}^M \psi_{m} g\left(\boldsymbol{\Theta}^{(m)}; t_1\right)^{-1} \boldsymbol{\pi}^{(m)} \mathbb{Q}_{1}^{(m)} \mathbb{Q}_{2}^{(m)}\ldots \mathbb{Q}_{K}^{(m)}\left\{ \mathbb{Q}_{D}^{(m)}\right\} ^{\delta}\mathbf{1}.
\end{eqnarray}
Likelihoods (\ref{eq:Lh0}) or (\ref{eq:Lh}) accommodate heterogeneities in the latent process even if no information regarding individuals' disease type is available.

\subsection{Incorporating auxiliary information on disease types}
\label{sec:method_3}

The likelihoods (\ref{eq:Lh0}) and (\ref{eq:Lh}) are constructed by summing up the mixture components corresponding to different disease types and associated possible underlying paths related to the observation process $Y(t)$.
In some cases, auxiliary information may help ascertain the disease type or the occupancy of certain states in the underlying process. 
For strictly progressive models, if it is known that a state of the underlying process has been occupied during the study, the participant may stay in this state or progress to the subsequent "later" states by the end of the observation period. 
In contrast, if a state is known to never have been occupied, the participant's path must end in the "earlier" states before that state.

For example, in the Nun Study as depicted in Figure~\ref{fig:model}, participants consented to brain donation after death, and postmortem neuropathological assessments through autopsy determined the presence and severity of AD pathology. 
For simplicity, we will use "with AD pathology" to represent the presence of AD pathology sufficient to meet diagnostic criteria for AD.
Recall that disease type I corresponds to non-AD dementia, while type II corresponds to AD.
For an autopsied participant, we may observe one of the following cases: 
(i) For those with both dementia and AD pathology, they are classified as type II, and they end up in the state of death with dementia and AD pathology ($Z^{(2)}$, state 6).
(ii) For those with AD pathology but not dementia, they are also classified as type II, and they end up in the state of death with AD pathology ($Z^{(2)}(t)$, state 5).
(iii) For those with dementia but not AD pathology, they are classified as type I, and they end up in the state of death with dementia ($Z^{(1)}(t)$, state 4).
(iv) Finally, for those with neither dementia nor AD pathology, the participant either belongs to type I, or is susceptible to type II but has died prior to the development of AD pathology. 
In this case, the participant may end up in state 3 of $Z^{(1)}$ or state 4 of $Z^{(2)}$.
When observations are intermittent, participants may develop dementia between the last visit and death. 
Therefore, for individuals who were not observed in the dementia state, we need to consider both (i) and (ii) if they have AD pathology, or (iii) and (iv) if they did not have AD pathology.
 
For strictly progressive disease processes, where all states have unique accessibility, a general approach to incorporate auxiliary information about the underlying processes can be outlined.
For an underlying process $Z^{(m)}(t)$ with the state set $\mathcal{J}^{(m)}$, define $\mathcal{A}^{(m)}$ as a collection of states where the participant could possibly end up, given the auxiliary information. 
We then define a vector $\mathbf{r}^{(m)}$ with the $j$-th element $r_j^{(m)}=I(j \in\mathcal{A}^{(m)}),  j\in \mathcal{J}^{(m)}$.
The likelihood of incorporating the auxiliary information is then expressed as:
\begin{eqnarray}
\label{eq:Lh-aux}
\mathcal{L} =\sum_{m=1}^M \psi_{m} g\left(\boldsymbol{\Theta}^{(m)}; t_1\right)^{-1} \boldsymbol{\pi}^{(m)} \mathbb{Q}_{1}^{(m)} \mathbb{Q}_{2}^{(m)}\ldots \mathbb{Q}_{K}^{(m)}\left\{ \mathbb{Q}_{D}^{(m)}\right\} ^{\delta} \mathbf{r}^{(m)}\;, 
\end{eqnarray}
and multiplication by vector $\mathbf{r}^{(m)}$ serves the purpose of summing over possible paths that are consistent with the auxiliary disease type and state occupancy information. 

Returning to the Nun Study application, for individuals classified to have disease type II due to AD pathology (cases (i) and (ii)), the possible states at the end of observation are $\mathcal{A}^{(1)}=\emptyset$ and $\mathcal{A}^{(2)}= \{ 2, 3, 5, 6\}$, with corresponding indicator vectors $\mathbf{r}^{(1)}=(0, 0, 0, 0)^T$ and $\mathbf{r}^{(2)}=(0, 1, 1, 0, 1, 1)^T$.
Individuals without AD pathology are either type I individuals or type II individuals who have not reached the pathology state yet by the last observation. 
This implies $\mathcal{A}^{(1)}=\mathcal{J}^{(1)}$, $\mathcal{A}^{(2)}= \{1, 4\}$, $\mathbf{r}^{(1)}=(1,1,1,1)^T$, and $\mathbf{r}^{(2)}=(1, 0, 0, 1, 0, 0)^T$, corresponding to cases (iii) and (iv).
For individuals observed in the dementia state, the likelihood will only consider paths ending in death with dementia states.
Given that neuropathological assessments are postmortem, neuropathology data are unavailable for living individuals or those where full postmortem assessment was impossible. 
In this case, $\mathcal{A}^{(m)}=\mathcal{J}^{(m)}$ and $\mathbf{r}^{(m)}=\mathbf{1}_{J^{(m)}}$, $m=1,2$, reducing (\ref{eq:Lh-aux}) to (\ref{eq:Lh}).

\section{Simulation Studies}
\label{sec:sim}

\subsection{Performance of MHMM}
To assess the performance of our proposed MHMM, we conducted simulation studies based on the disease processes illustrated in Figure~\ref{fig:model}, mimicking the Nun Study scenario.
We defined the study period from time 0 to time 1, with observations of $Y(t)$ recorded at equally spaced intervals: ${0, 0.25, 0.5, 0.75, 1}$.
Assuming a 50\% chance for each disease type, the initial state probabilities for the two disease types are $\boldsymbol{\pi}^{(1)}=(0.7,0.3,0)$ and $\boldsymbol{\pi}^{(2)}=(0.4,0.3,0.3,0,0)$ respectively. 
Transition intensities were determined to achieve an overall right censoring rate of 20\%, and solved by setting $\lambda_{12}^{(1)}/\lambda_{13}^{(1)}=2$, $\lambda_{24}^{(1)}/\lambda_{13}^{(1)}=1.5$ for disease type I, and $\lambda_{12}^{(2)}/\lambda_{14}^{(2)}=2.2$, $\lambda_{25}^{(2)}/\lambda_{14}^{(2)}=1.8$, $\lambda_{36}^{(2)}/\lambda_{14}^{(2)}=2.5$ for disease type II. 
Participants were assumed to enter the study at time 0, with no delayed entry.
For individuals whose death occurred during the study period, we assumed that their disease type becomes known with a probability of 80\%. 
This reflects the scenario where postmortem assessments reveal the underlying disease pathology of deceased participants.
On the other hand, for individuals censored at the end of the follow-up period, we treated their disease type as unknown. 
This mirrors situations where participants remain alive until the end of data collection, without further information available about their disease type.

Table~\ref{tb:5} gives the simulation results based on 700 replications and a sample size of 500. 
The maximization procedure was conducted under the constraint that the ratio $\pi_2^{(2)}/\pi_1^{(2)}$ is known to avoid identifiability and estimability issues. 
More discussion around these issues is given in Section~\ref{sec:iden}. 
The simulation results indicate that the average estimates are close to the true values of the parameters.
The standard errors are the empirical standard errors of the estimates from 700 replications. 
For each replication, 95\% confidence intervals (CIs) were calculated using the estimates and model-based standard errors. 
The coverage probabilities of the CIs are close to 0.95 for most cases, suggesting that the estimated standard deviations are reasonable.

\begin{table}[h!]
	\caption{Simulation results based on 700 replications with sample size 500. The MLEs are obtained while assuming known $\pi_2^{(2)}/\pi_1^{(2)}$. Est: average estimates; SE$_{\textrm{emp}}$: empirical standard errors; CI Coverage: coverage probability for 95\% confidence intervals.   }	
	\label{tb:5}
	\centering \begin{tabular}{ccccc}
		\hline \\[-1em]
		& True & Est & SE$_{\textrm{emp}}$ & CI Coverage \\   \hline 
		\\[-1em]
		\multicolumn{5}{c}{\it{Disease  Type I}}\\
		\\[-1em]
		$\lambda^{(1)}_{12}$  & 2.383 &2.984  &0.669 &97.348\\    
		$\lambda^{(1)}_{13}$  & 1.191 &0.754  &0.514 &99.242 \\    
		$\lambda^{(1)}_{24}$ & 1.787 &1.838  &0.299 &92.614\\       
		\\[-1em]
		\multicolumn{5}{c}{\it{Disease  Type II}}\\
		\\[-1em]
		$\lambda^{(2)}_{12}$ & 1.802 &2.202  &0.728 &99.432\\   
		$\lambda^{(2)}_{14}$ & 0.819 &0.745 &0.645 &94.886\\   
		$\lambda^{(2)}_{23}$ & 2.457 &2.251  &0.515 &94.129\\    
		$\lambda^{(2)}_{25}$ & 1.474 &2.135  &0.453 &87.500\\    
		$\lambda^{(2)}_{36}$ & 2.047 &2.081  &0.364 &91.667\\  
		\\[-1em]
		\multicolumn{5}{c}{\it{Other Parameters}}\\
		\\[-1em]
		$\psi_1$  & 0.500 &0.443 &0.048 &94.886\\ 
		$\pi^{(1)}_1$  & 0.700 &0.661 &0.045 &98.295\\
		$\pi^{(2)}_1$  & 0.400 &0.416 &0.020 &93.561\\ 
		\\[-1em]
		\hline \end{tabular}	
\end{table}

\subsection{Identifiability and Estimability}
\label{sec:iden}

The identifiability issue arises when multiple sets of parameters lead to the same observed process. This problem is common in HMMs due to the incomplete nature of the observation scheme and the potential complexity of the models, as noted by \cite{titman2010}. 
In Appendix A, we provide an example demonstrating the nontrivial nature, if not intractability, of identifiability issues. 
Even the simplest models may suffer from this problem. 
Given that the proposed MHMM has a more complicated structure, the non-identifiability issue is more likely to occur.
We demonstrate a specific scenario for the model given in Figure~\ref{fig:model} in Appendix B, highlighting the potential identifiability challenge in our analysis.

Even if all parameters are identifiable, difficulties in estimation may arise due to the complex model structure involving mixed populations and hidden states, along with intermittently observed data. 
In such cases, local maxima in the likelihood function may exist, leading the maximizing procedure to converge to a suboptimal solution.
Furthermore, "weak identifiability" can be a challenge, where the likelihood function has flat regions around the maximum likelihood value. 
This phenomenon, commonly observed in latent class models, results in multiple different but close estimates for the maximum likelihood.

Addressing the identifiability and estimability issues often involves increasing the amount of available information or reducing the number of unknown parameters by introducing constraints. 
In our simulation study for the model depicted in Figure~\ref{fig:model}, we employed a dimension reduction approach to mitigate these issues by imposing constraints on certain parameters. 
In Appendix C, we explored several options of constraints through simulation studies. 
The results showed why the constraint on the ratio of being in the pathology state versus dementia-free for disease type II, i.e., $\pi^{(2)}_2/\pi^{(2)}_1$, is chosen. 
However, the choice of constraint depends on the structure of the model and the values of the true intensities. 
Developing standard solutions that apply to general identifiability and estimability issues in MHMMs remains a challenging problem.

Incorporating additional constraints derived from external sources or previous studies can provide valuable insights and improve the estimation process in complex models.
For instance, if there are known relationships between transition intensities or differences in disease progression rates between different groups, such information can be directly incorporated as constraints in the model. 
This can help ensure that the estimated parameters are consistent with the existing knowledge base and can lead to more reliable inferences.

Incorporating previous understanding of the disease process in a Bayesian framework can also be a powerful approach.
In Bayesian inference, prior information or beliefs about the parameters can be formally incorporated into the analysis through prior distributions. 
These priors can reflect existing knowledge about the disease process, such as information on transition intensities, disease progression rates, or other relevant factors obtained from previous studies or expert opinions.
The Bayesian approach not only helps with the identifiability/estimability issues but also speeds up the numerical calculation. 
We will explore this approach using the Nun Study in Section~\ref{sec:Nunbayes}.

\section{Application to the Nun Study: AD vs. non-AD dementia}
\label{sec:resultsNun}

\subsection{Frequentist Analysis}

Table~\ref{tb:3} presents the MLE of the parameters obtained by directly maximizing the likelihood (\ref{eq:Lh-aux}). 
To address identifiability issues inherent in hidden Markov models, the probability of being in the AD pathology pre-dementia state at the initial age of 75 was fixed at zero, while being in the dementia state following the onset of AD pathology was allowed at baseline.  
The 95\% CIs for the logarithm of the transition intensities were calculated using Wald-statistic and model-based standard errors.  
The proposed MHMM is compared to a time-homogeneous three-state model without any hidden structure (the naive approach).
According to Table~\ref{tb:3}, for individuals with dementia in the study population, the probability of being classified as AD is estimated to be 0.738, with a 95\% CI of (0.526, 0.877). 
This estimate closely aligns with results from the Aging, Demographics, and Memory Study (ADAMS), the first population-based study of dementia in the United States, which reported a prevalence of AD and all-cause dementia among females of all ages as 11.48\% and 15.74\%, respectively, yielding a ratio of 0.729 \citep{plassman2007prevalence}. 
Additionally, the intensity of progressing to the dementia state following the onset of AD pathology is estimated to be more than twice as high compared to non-AD dementia, i.e., $\widehat{\lambda}^{(2)}_{23}=0.122$ vs. $\widehat{\lambda}^{(1)}_{12}=0.055$. 
It takes an estimated average of 6.536 years, starting from the dementia-free state, for a type II individual to develop AD pathology, and then a further 8.197 years to proceed to dementia, compared to 18.182 years for individuals of type I to develop non-AD dementia.
Furthermore, compared to dementia-free individuals, the development of dementia increases the risk of death by 4.284 times for non-AD dementia and 7.697 times for AD.

\begin{table}[t!]
	\caption{Estimated transition intensities for dementia from the Nun Study data using MHMM. Parameters are estimated while fixing $\pi^{(2)}_{2}=0$. 
 Confidence intervals (CI) are obtained by taking exponentials of the CI bounds for the log transition intensities, which are calculated based on the MLEs and model-based standard errors for the log transition intensities.}
	
	\label{tb:3}
	\centering \begin{tabular}{llccc}
		\hline\\[-1em]
		
		&   & Est  & 95\% CI\\  \\[-1em] \hline
		\\[-1em]
		\multicolumn{4}{c}{\it{Disease  Type I}}\\
		\\[-1em]
		Dementia-free to Dementia &$\lambda^{(1)}_{12}$ & 0.055 & (0.024, 0.130)\\   
		Dementia-free to Death &$\lambda^{(1)}_{13}$& 0.088 &  (0.050, 0.154) \\  
		Dementia to Death &$\lambda^{(1)}_{24}$& 0.377 &  (0.290, 0.489) \\    \\[-1em]
		\multicolumn{4}{c}{\it{Disease  Type II}}\\
		\\[-1em]
		Dementia-free to Pathology &$\lambda^{(2)}_{12}$& 0.153  &(0.075, 0.311) \\  
		Dementia-free to Death &$\lambda^{(2)}_{14}$& 0.033  &(0.004, 0.288)\\  
		Pathology to Dementia&$\lambda^{(2)}_{23}$& 0.122  &(0.094, 0.158)\\    
		Pathology to Death &$\lambda^{(2)}_{25}$& 0.088  &(0.066, 0.117) \\      
		Dementia to Death &$\lambda^{(2)}_{36}$&0.254  &(0.223, 0.289) \\ 
		\\[-1em]
		\multicolumn{4}{c}{\it{Other Parameters}}\\
		\\[-1em]
		Mixture Probability of AD & $\psi_2$  & 0.738  &  (0.526, 0.877)\\ 
		Initial Probability of $Z^{(1)}(t)$& $\pi^{(1)}_{1}$  & 0.977  &(0.028, 1.000) \\ 
		Initial Probability of $Z^{(2)}(t)$&$\pi^{(2)}_{1}$  & 0.974  &(0.825, 0.997)\\ \\[-1em]
		\hline 
	\end{tabular}

\end{table}

Using the estimated transition intensities and initial probabilities, one can gain insights into the disease process by estimating several useful quantities.
For example, prevalence is the proportion of individuals with dementia among living people of a given age, while cumulative incidence represents the probability of a specific type of dementia by age $t$. 
Details of the calculation for the Nun Study model, as shown in Figure~\ref{fig:model}, are provided in Appendix D.
Figure~\ref{fig:prevFreq} illustrates the estimated all-cause prevalence, type-specific prevalence, and type-specific cumulative incidence for the Nun Study cohort aged 75 to 100.  
The left plot compares the estimated all-cause prevalence obtained from the proposed MHMM (blue solid curve) with the one from a naive method using a time-homogeneous illness-death model (red dashed curve).
The MHMM results align well with the population prevalence for women aged 70--79, 80--89, and 90+ as reported in the ADAMS study \citep{plassman2007prevalence}. 
On the other hand, the naive approach, which does not differentiate between dementia subtypes, overestimates the prevalence for the youngest group while underestimating it for the oldest group.
The middle panel demonstrates that the prevalence of AD rises dramatically with age, while the prevalence of non-AD dementia exhibits less significant variation with age (right-hand panel). 
Consequently, the proportion of AD among all-cause dementia increases with age. 
The cumulative incidence trends differ between the two disease types, with AD showing slower increases at younger ages but more rapid rises at older ages compared to non-AD dementia.

\begin{figure}[t!]
	\centerline{\includegraphics[width=100ex]{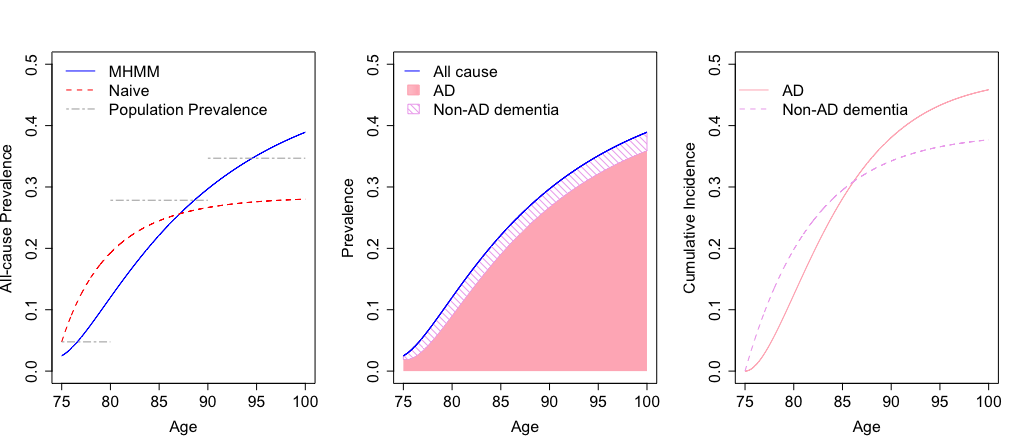}}
	\caption{Estimated prevalence and cumulative incidence from MHMM. 
 (1) Left plot: All-cause dementia prevalence from a naive three-state time-homogeneous model (red dashed curve) and the proposed MHMM (blue solid curve). 
 Dashed horizontal lines are the population prevalence for women aged 70--79, 80--89, and 90+ reported from the Aging, Demographics, and Memory Study (ADAMS).
 (2) Middle plot: All-cause and type-specific prevalence from MHMM. 
 The blue curve represents dementia prevalence from both types.
 Heights of the red and the purple shaded regions are prevalences of AD and non-AD dementia respectively. 
 (3) Right plot: Type-specific cumulative incidence for AD (red solid curve) and non-AD dementia (purple dashed curve).}
	\label{fig:prevFreq} 
\end{figure}

\subsection{Bayesian Analysis}
\label{sec:Nunbayes}

The Bayesian method has become a popular tool for making statistical inferences using complex models due to the convenience of numerical calculation and the ability to incorporate knowledge of the parameters. 
It usually involves the assignment of prior distributions to the parameters and, possibly, a set of hyperparameters of the prior distributions. 
Bayesian inference is then made based on the posterior distributions of the parameters, which are derived based on the priors and the conditional distribution of the observed data given the values of the parameters.
The standard point estimate of a parameter is the posterior mean.

\begin{figure}[h!]
	\centerline{\includegraphics[width=65ex]{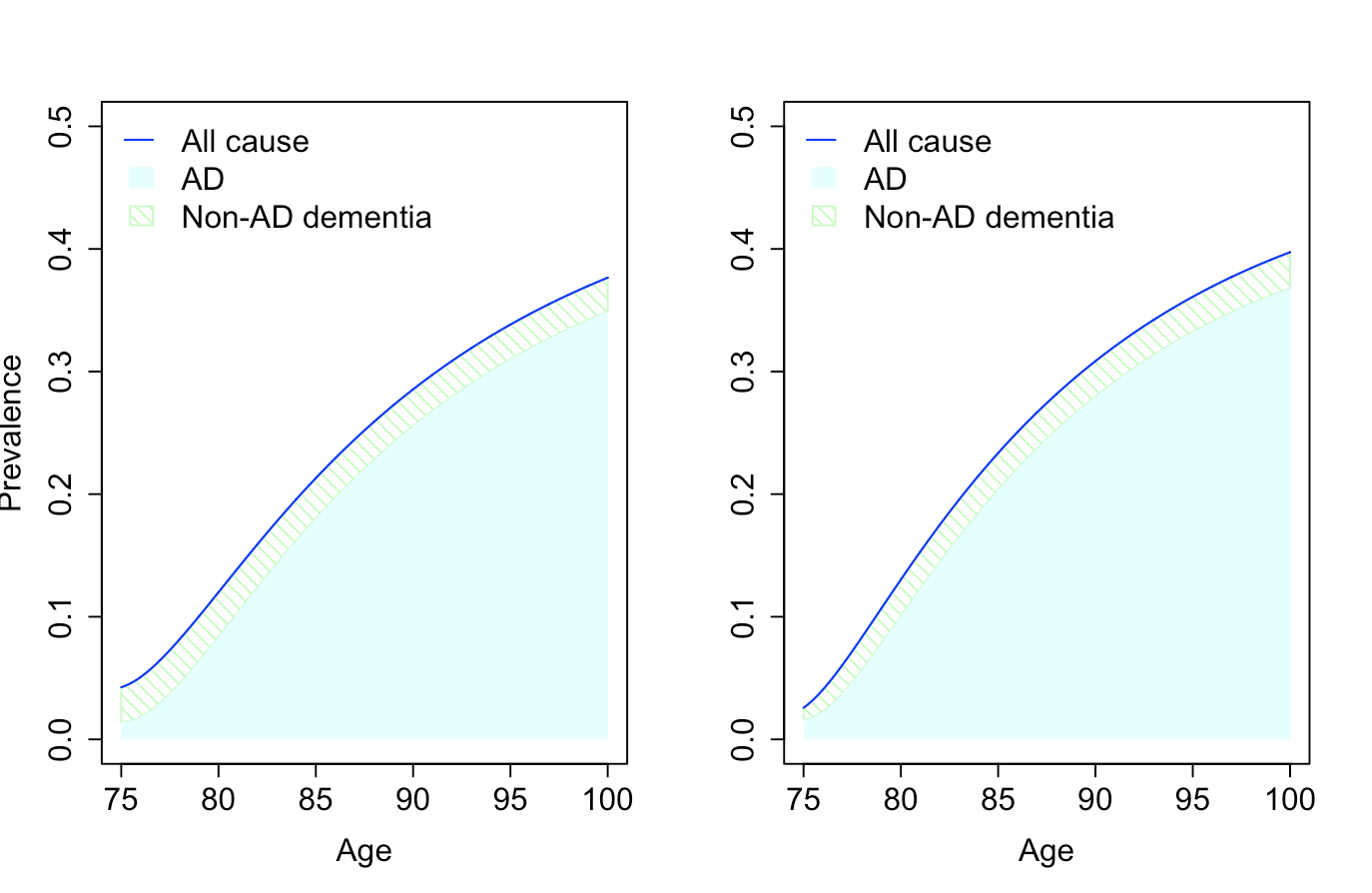}}
 \caption{Comparison of prevalence using a non-informative prior (left plot) and an informative prior based on ADAMS (right plot). 
 The blue curves represent all-cause prevalence.
 Heights of the blue and the green shaded regions are the prevalence of AD and non-AD dementia respectively. 
 ADAMS: Aging, Demographics, and Memory Study \citep{plassman2007prevalence}.}
	\label{fig:prevBayes} 
\end{figure}

As mentioned previously, the ADAMS study \citep{plassman2007prevalence} provides valuable population-level prevalence data on AD and dementia. 
In our analysis, we incorporated this information into the fitting of the proposed MHMM by using it as prior knowledge.
Specifically, we used the ratio of the prevalence of AD vs non-AD dementia from ADAMS \citep{plassman2007prevalence}, 0.729, as the mean of the prior distribution of mixture probability $\psi_2$. 
Additionally, we utilized the population prevalence of non-AD dementia and AD for the age group 71--79 as the means of the prior distributions for the initial probabilities $\pi_2^{(1)}$ and $\pi_3^{(2)}$, which are 2.46\% and 2.30\% respectively.
The numerical calculations used the Gibbs Sampler with 2000 iterations, conducted by the R function ``jags()"  in the $R2jags$ package \citep{R2jags}. 
Convergence of the Gibbs Sampler is evaluated by the effective number of iterations $N_{\textrm{eff}}$ and the potential scale reduction factor $\widehat{R}$ \citep{gentleman1994}. 
Usually, we expect $N_{\textrm{eff}}\geq30$ and $\widehat{R}\approx1$ if the algorithm converges successfully. 
Further details about the prior distributions and the estimated transition intensities are provided in Appendix E.
Comparing the results, it is evident that using informative priors leads to narrower Bayesian CIs.
Figure~\ref{fig:prevBayes} illustrates the prevalence estimated from these Bayesian estimates.
The Bayesian estimation utilizing informative priors yields estimates similar to those obtained through the frequentist approach (the middle plot of Figure~\ref{fig:prevFreq}). 
Importantly, no identifiability or estimability issues were encountered during the Bayesian estimation process.

\section{Discussion}
\label{sec:remarks}

In this paper, we proposed an MHMM, which provides the ability to model non-Markovian transition times while utilizing the nice properties of Markov models. 
This approach proves especially valuable in scenarios where the underlying disease exhibits multiple subtypes and where relaxing the Markov assumption on the observation model is necessary. 
Furthermore, our method allows for the integration of potential information on hidden state occupancy, thereby enhancing both efficiency and estimability.

An advantage of MHMMs lies in their ability to provide interpretable estimates for cohorts exhibiting either a singular disease type or a mixture of disease types. 
In our investigation of the dementia disease process, the model could effectively capture variations in prevalence between AD and non-AD dementia. 
However, one needs to be careful when relating estimates from mixed-type cohorts to specific disease types due to potential challenges with identifiability and estimability. 
While theoretically the model could accommodate additional disease types, there exists a trade-off between model complexity and estimation performance. 

We have shown that even the simplest HMM suffers from identifiability and estimability issues. 
This suggests that strategies to identify and contain these issues are essential to obtaining good estimates from the more complicated MHMM. 
We employed a combination of analytical and empirical methods to detect and address these issues, in particular by imposing constraints to reduce the dimension of unknown parameters or by introducing prior information through Bayesian approaches. 
It is worth noting that the selection of parameters for informative priors should be tailored to the model structure and the characteristics of the dataset. 
The specific procedures for implementing these strategies may vary depending on the dataset and the particular disease process under investigation.

Integrating risk factors into the transition intensities of the model opens avenues for a more comprehensive understanding of disease progression.
Let $\mathbf{X}=\left(X_1,X_2,\ldots,X_P\right)$ be the vector of the risk factors.
For underlying disease type $m$, let $\mathbf{\beta}^{(m)}_{ij}=(\beta^{(m)}_{ij,1},\beta^{(m)}_{ij,2},\ldots,\beta^{(m)}_{ij,P})$ be the log instantaneous rate ratios for transition from state $i$ to state $j$. 
The covariate effects can be modeled through a proportional hazard model $\lambda^{(m)}_{ij}\left(t;\mathbf{X}\right) = \lambda^{(m)}_{ij0}(t)e^{\mathbf{X}^{\textrm{T}}\mathbf{\beta}_{ij}^{(m)}}$.
This assumption of universal effects may help reduce the parameter dimension and mitigate identifiability and estimability issues. 
Furthermore, if the states correspond to disease stages or have other meaningful interpretations, auxiliary information on these parameters or ratios may be obtained from existing literature.  

Modeling the missingness mechanism of the state occupancy data is another extension, particularly in scenarios when it depends on disease stages. 
In the Nun Study, the timing of postmortem neuropathological assessments was driven by the timing of participant deaths, suggesting that missingness in neuropathology data may be completely at random. 
However, verifying this assumption can be challenging with panel observations for a multistate process.
In cases where missingness depends on disease stages, a joint modeling approach that incorporates both the transition intensities and the mixture components is necessary. 
This approach allows for the simultaneous consideration of disease progression and missingness mechanisms, offering a more comprehensive understanding of the underlying process. 

In conclusion, the MHMM for intermittently observed multistate processes presents a powerful framework for detecting heterogeneity and relaxing the Markov assumptions inherent in traditional models. 
By incorporating auxiliary information and addressing identifiability and estimability issues, this approach offers enhanced flexibility and accuracy in modeling complex disease dynamics.
We anticipate that this work will stimulate further interest and investigations in both methodological research and application to diverse domains.

\section*{Acknowledgements}
This work was supported by the Natural Science and Engineering Research
Council of Canada through Discovery Grant RGPIN 03648 (Dr. Leilei Zeng) and RGPIN 03688 (Dr. Mary E. Thompson) and the Canadian Institutes of Health Research through grant MOP 137035 (Dr. Suzanne L. Tyas).

\bibliographystyle{imsart-nameyear} 
\bibliography{main-bib.bib}

\newpage
\appendix

\section{A simple example of non-identifiability in hidden Markov models (HMM)}

\begin{figure}[h!]
		\centerline{\includegraphics[width=.5\textwidth]{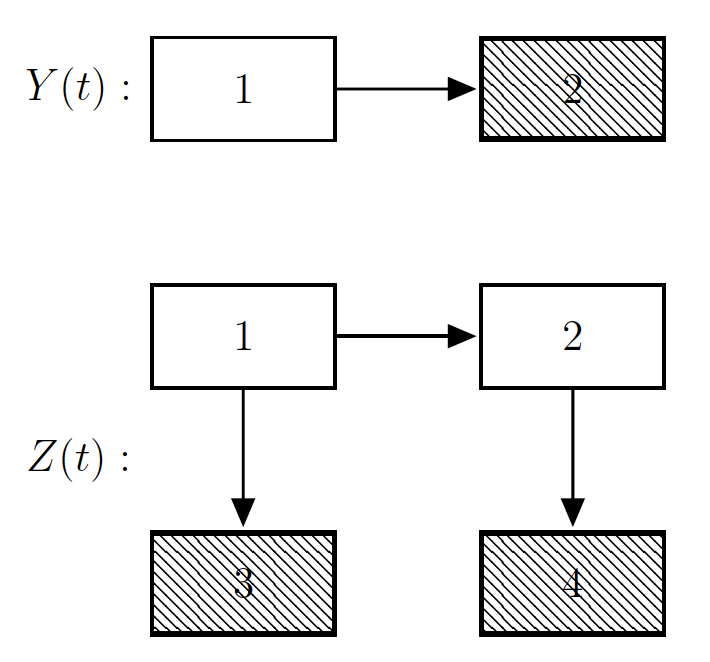}}
	\caption{A simple example of HMM. The underlying states and their corresponding observed states share the same filling pattern.}
	\label{fig:A1}
\end{figure}

To show that the identifiability is highly nontrivial, if not intractable, we explore the issue starting from a simple HMM as in Figure \ref{fig:A1}, where the observation process $Y(t)$ contains only one transient state and one absorbing state, and the underlying model $Z(t)$ has four states.
The unknown parameters involved in such a model are $\mathbf{\theta}=\left(\pi_1,\lambda_{12},\lambda_{13},\lambda_{23}\right)$; however, it can be shown that the likelihood is determined by three effective parameters.  
More specifically, for every $0<\rho<1/\pi_1$, parameter set $\left(\rho\pi_{1}, \lambda_{12},\lambda_{13}-(1-\rho)\lambda_{23}/\rho,\lambda_{23}\right)$ gives the same likelihood value.

\section{An example of non-identifiability in the proposed MHMM}

The closed-form likelihood for a disease process model as shown in Figure~\ref{fig:model} is extremely complicated.
Therefore, we simplify the likelihood by restricting the possible path of each individual.
Consider the case where all of the participants were followed from time 0 until their deaths. 
Assume that all of the participants started in the disease-free state and were diagnosed with dementia before they died, and that postmortem neuropathological information is missing for everyone.
Let $a_1$ and $a_2$ be the time at the first and the last assessment time that the participant is in the disease-free state, i.e. $a_1=\{t_k;Y(t_k)=1\}$, $a_2=\max\{t_k;Y(t_k)=1\}$, $k=0,1,\ldots,K$. 
Similarly, let $a_3=\min\{t_k;Y(t_k)=2\}$, $a_4=\max\{t_k;Y(t_k)=2\}$, and $a_5=t_D$. 
Denote $a_{uv}=a_v-a_u$ for $u<v, u,v\in\{1,2,\ldots,5\}$. 
If $\lambda^{(1)}_{12}+\lambda^{(1)}_{13}\neq \lambda^{(1)}_{23}$ and $\lambda^{(2)}_{12}+\lambda^{(2)}_{14}\neq \lambda^{(2)}_{23}+\lambda^{(2)}_{25}\neq \lambda^{(2)}_{35}$, the closed-form for likelihood (2) of Shi, et al. (2024) is
\begin{eqnarray}
\mathcal{L}  & = & \psi_{1}\pi_{1}^{(1)}\lambda_{12}^{(1)}\lambda_{24}^{(1)}\frac{e^{-\left(\lambda_{12}^{(1)}+\lambda_{13}^{(1)}\right)a_{12}-\lambda_{24}^{(1)}\left(a_{23}+a_{35}\right)}-e^{-\left(\lambda_{12}^{(1)}+\lambda_{13}^{(1)}\right)\left(a_{12}+a_{23}\right)-\lambda_{24}^{(1)}a_{35}}}{\lambda_{12}^{(1)}+\lambda_{13}^{(1)}-\lambda_{24}^{(1)}}\nonumber\\
&  & +(1-\psi_{1})\left\{ \pi_{1}^{(2)}\lambda_{12}^{(2)}\lambda_{23}^{(2)}\lambda_{36}^{(2)}\left(\frac{e^{-\left(\lambda_{12}^{(2)}+\lambda_{14}^{(2)}\right)a_{12}-\lambda_{36}^{(2)}\left(a_{23}+a_{35}\right)}}{\left(\lambda_{12}^{(2)}+\lambda_{14}^{(2)}-\lambda_{36}^{(2)}\right)\left(\lambda_{23}^{(2)}+\lambda_{25}^{(2)}-\lambda_{36}^{(2)}\right)}\right.\right.\nonumber\\
&  & +\frac{e^{-\left(\lambda_{12}^{(2)}+\lambda_{14}^{(2)}\right)\left(a_{12}+a_{23}\right)-\lambda_{36}^{(2)}a_{35}}}{\left(\lambda_{12}^{(2)}+\lambda_{14}^{(2)}-\lambda_{23}^{(2)}-\lambda_{25}^{(2)}\right)\left(\lambda_{12}^{(2)}+\lambda_{13}^{(2)}-\lambda_{36}^{(2)}\right)} \nonumber\\
&  & \left.-\frac{e^{-\left(\lambda_{12}^{(2)}+\lambda_{14}^{(2)}\right)a_{12}-\lambda_{36}^{(2)}\left(a_{23}+a_{35}\right)}+e^{-\left(\lambda_{23}^{(2)}+\lambda_{25}^{(2)}\right)\left(a_{12}+a_{23}\right)-\lambda_{36}^{(2)}a_{35}}-e^{-\left(\lambda_{23}^{(2)}+\lambda_{25}^{(2)}\right)a_{12}-\lambda_{36}^{(2)}\left(a_{23}+a_{35}\right)}}{\left(\lambda_{12}^{(2)}+\lambda_{14}^{(2)}-\lambda_{23}^{(2)}-\lambda_{25}^{(2)}\right)\left(\lambda_{23}^{(2)}+\lambda_{25}^{(2)}-\lambda_{36}^{(2)}\right)}\right) \nonumber\\
&  & \left.-\pi_{2}^{(2)}\lambda_{23}^{(2)}\lambda_{36}^{(2)}\frac{e^{-\left(\lambda_{23}^{(2)}+\lambda_{25}^{(2)}\right)\left(a_{12}+a_{23}\right)-\lambda_{36}^{(2)}a_{35}}-e^{-\left(\lambda_{23}^{(2)}+\lambda_{25}^{(2)}\right)a_{12}-\lambda_{36}^{(2)}\left(a_{23}+a_{35}\right)}}{\lambda_{23}^{(2)}+\lambda_{25}^{(2)}-\lambda_{36}^{(2)}}\right\} ,\nonumber
\label{eq-likid}
\end{eqnarray}
where the vector of unknown parameters is 
\begin{eqnarray*}
	\mathbf{\theta}=\left(\begin{array}{ccccccccccccc}
		\psi_{1}, & \pi_{1}^{(1)}, & \lambda_{12}^{(1)}, & \lambda_{13}^{(1)}, & \lambda_{24}^{(1)}, & \pi_{1}^{(2)}, & \pi_{2}^{(2)}, & \lambda_{12}^{(2)}, & \lambda_{14}^{(2)}, & \lambda_{23}^{(2)}, & \lambda_{25}^{(2)}, & \lambda_{36}^{(2)}\end{array}\right).
\end{eqnarray*}
It is obvious that we are able to identify $\lambda_{12}^{(1)}+\lambda_{24}^{(1)}$, $\lambda_{24}^{(1)}$, $\lambda_{12}^{(2)}+\lambda_{14}^{(2)}$, $\lambda_{23}^{(2)}+\lambda_{25}^{(2)}$, $\lambda_{36}^{(2)}$, and $\psi_1$. 
Consider some $\rho_1\in (0,1/\pi^{(1)}_{1})$ and $\rho_2\in(0,1/(\pi^{(1)}_{2}+\pi^{(2)}_{2}))$; the following parameter vector gives the same likelihood as $\mathbf{\theta}$ does.
\begin{eqnarray*}
		\mathbf{\theta}^{*} & = & \left(\begin{array}{ccccc}
		\psi_{1}, & \rho_{1}\pi_{1}^{(1)}, & \frac{\lambda_{12}^{(1)}}{\rho_{1}}, & \left(1-\frac{1}{\rho_{1}}\right)\lambda_{12}^{(1)}+\lambda_{13}^{(1)}, & \lambda_{24}^{(1)},\end{array}\right. \nonumber\\
	&  & \left.\begin{array}{ccccccc}
		\rho_{2}\pi_{1}^{(2)}, & \rho_{2}\pi_{2}^{(2)}, & \lambda_{12}^{(2)}, & \lambda_{14}^{(2)}, & \frac{\lambda_{23}^{(2)}}{\rho_{2}}, & \left(1-\frac{1}{\rho_{2}}\right)\lambda_{23}^{(2)}+\lambda_{25}^{(2)}, & \lambda_{36}^{(2)}\end{array}\right)
\end{eqnarray*}
Therefore, there is an identifiability issue in this case.

\section{Supplementary simulations on the choice of constraints}

Table \ref{tb:6} gives several options of constraints and their performance based on the same simulation setting as in Section 4.
These scenarios include $S_0$ - directly maximizing the likelihood without constraint; $S_1$ - assuming that the ratio of being in the pathology state versus dementia-free for disease type II, i.e., $\pi^{(2)}_2/\pi^{(2)}_1$, is known; $S_2$ - controlling the ratio of intensities from dementia-free to death of the two disease types; $S_3$ - controlling the previous two ($S_1$ and $S_2$) at the same time; and $S_4$ - controlling the ratio of intensities from dementia to death of the two disease types. 
The estimates and the model-based standard errors in Table \ref{tb:6} are averaged over 100 replications. 
Scenarios $S_0$, $S_2$ and $S_4$ do not impose constraints on the initial probabilities and the MLE $\lambda_{12}^{(2)}$ yielded zero values, suggesting maxima outside the boundary of the parameter space. 
Scenarios $S_1$ and $S_3$ produced valid estimates for $\lambda_{12}^{(2)}$, while $S_1$ required less reduction in dimension of the parameter. 
This is why scenario $S_1$ was chosen.

\begin{table}[h!]
	\caption{Comparison of the choices of constraints based on simulation studies with 700 replications and sample size 500}
	\label{tb:6} 
	\centering
	\begin{tabular}{cccccccccccc}
		\hline \\[-1em]
		\multicolumn{2}{c}{Constraints} & \multicolumn{2}{c}{$S_0$: No Cons.} & \multicolumn{2}{c}{$S_1$:  $\frac{\widehat{\pi}_{2}^{(2)}}{\widehat{\pi}_{1}^{(2)}}=\frac{\pi_{2}^{(2)}}{\pi_{1}^{(2)}}$}& \multicolumn{2}{c}{$S_2$:  $\frac{\widehat{\lambda}_{14}^{(2)}}{\widehat{\lambda}_{13}^{(1)}}=\frac{\lambda_{14}^{(2)}}{\lambda_{13}^{(1)}}$}&\multicolumn{2}{c}{$S_3$: $S_1$\&$S_2$}& \multicolumn{2}{c}{$S_4$:  $\frac{\widehat{\lambda}_{36}^{(2)}}{\widehat{\lambda}_{14}^{(2)}}=\frac{\lambda_{36}^{(2)}}{\lambda_{14}^{(2)}}$} \\ \hline
		\\[-1em] 
		& True & Est & SE & Est & SE & Est & SE & Est & SE & Est & SE\\ \\[-1em]
		\hline
		\\[-1em]
		& & \multicolumn{10}{c}{\it{Transition Intensities of $Z^{(1)}(t)$}}\\
		\\[-1em]
		$\lambda_{12}^{(1)}$ & 2.383 & 2.441 & 0.713 & 3.166 & 1.029 & 2.269 & 0.356 & 3.595 & 0.812 & 2.832 & 0.599 \\ 
		$\lambda_{13}^{(1)}$ & 1.191 & 1.272 & 0.275 & 0.947 & 0.555 & 1.275 & 0.257 & 0.640 & 0.382 & 1.214 & 0.319 \\ 
		$\lambda_{24}^{(1)}$ & 1.787 & 2.157 & 0.323 & 2.169 & 0.323 & 2.148 & 0.320 & 2.160 & 0.321 & 2.179 & 0.325 \\ 
		\\[-1em]
		& & \multicolumn{10}{c}{\it{Transition Intensities of $Z^{(2)}(t)$}}\\
		\\[-1em]
		$\lambda_{12}^{(2)}$& 1.802 & 0.000 & 0.005 & 1.447 & 0.424 & 0.000 & 0.000 & 1.267 & 0.204 & 0.000 & 0.004 \\ 
		$\lambda_{14}^{(2)}$ & 0.819 & 0.274 & 0.835 & 0.610 & 0.466 & 0.877 &0.177  & 0.440 &0.263  & 0.803 & 0.127 \\ 
		$\lambda_{23}^{(2)}$ & 2.457 & 1.386 & 0.254 & 2.474 & 0.610 & 1.390 & 0.256 & 2.296 & 0.480 & 1.346 & 0.245 \\ 
		$\lambda_{25}^{(2)}$& 1.474 & 1.555 & 0.290 & 2.665 & 0.674 & 1.577 & 0.283 & 2.455 & 0.503 &1.445  & 0.229 \\ 
		$\lambda_{36}^{(2)}$ & 2.047 & 1.449 & 0.226 & 1.437 & 0.221 & 1.454 & 0.227 & 1.438 & 0.221 & 1.435 & 0.220 \\ 
		\\[-1em]
		& & \multicolumn{10}{c}{\it{Other Parameters}}\\
		\\[-1em]
		$\psi_1$ & 0.500 & 0.495 & 0.161 & 0.437 & 0.153 & 0.508 & 0.103 & 0.409 & 0.087 & 0.466 & 0.102 \\ 
		$\pi_1^{(1)}$ & 0.700 & 0.669 & 0.312 & 0.625 & 0.329 & 0.677 & 0.249 & 0.599 & 0.212 & 0.649 & 0.243 \\ 
		$\pi_1^{(2)}$ & 0.400 & 0.076 & 0.076 & 0.419 & 0.051 & 0.057 & 0.027 & 0.426 & 0.039 & 0.111 & 0.058 \\ 
		$\pi_2^{(2)}$ & 0.300 & 0.627 & 0.311 & 0.314 & 0.038 & 0.639 & 0.234 & 0.320  & 0.029 & 0.607 & 0.245 \\

		\\[-1em]
		\hline
	\end{tabular}
	
\end{table}

\section{Calculation of prevalence and cumulative incidence}

Prevalence represents the proportion of individuals being in the disease state among the alive population of a given age. 
Using the proposed MHMM model, we can calculate both all-cause and type-specific prevalence of dementia.
Let $\mathbb{P}^{(m)}(s,s+t)=\exp(t\mathbb{A}^{(m)})$ be the probability matrix of disease type $m$ and $P_{jk}^{(m)}$ be its $(j, k)$-th element.
The prevalence of non-AD dementia (type I disease) at age $t$ is
\begin{eqnarray}
\label{e:prev1}
& P(\Delta=1,Y(t)=2\mid Y(t)<3) \nonumber \\
 = & \frac{\psi_{1}\sum_{j=1}^{2}\text{\ensuremath{\pi}}_{j}^{(1)}P_{j2}^{(1)}(t)}{\psi_{1}\sum_{j=1}^{2}\pi_{j}^{(1)}(P_{j1}^{(1)}(t)+P_{j2}^{(1)}(t))+\psi_{2}\sum_{j=1}^{3}\pi_{j}^{(2)}(P_{j1}^{(2)}(t)+P_{j2}^{(2)}(t)+P_{j3}^{(2)}(t))}.
\end{eqnarray}
Similarly, the prevalence of AD (type II disease) at age $t$ is
\begin{eqnarray}
\label{e:prev2}
& P(\Delta=2,Y(t)=2\mid Y(t)<3) \nonumber \\
 = & \frac{\psi_{2}\sum_{j=1}^{3}\pi_{j}^{(2)}P_{j3}^{(2)}(t)}{\psi_{1}\sum_{j=1}^{2}\pi_{j}^{(1)}(P_{j1}^{(1)}(t)+P_{j2}^{(1)}(t))+\psi_{2}\sum_{j=1}^{3}\pi_{j}^{(2)}(P_{j1}^{(2)}(t)+P_{j2}^{(2)}(t)+P_{j3}^{(2)}(t))}.
\end{eqnarray}
The all-cause prevalence at age $t$ is $P(Y(t)=2)/P(Y(t)<3)$, which can be estimated by the summation of (\ref{e:prev1}) and (\ref{e:prev2}).

Another widely used quantity in time-to-event analysis is cumulative incidence, which represents the probability of failure by a given age.
Using strictly progressive underlying models helps with the calculation of cumulative incidence.  
In the Nun Study case, for the type I disease process, we define the cumulative incidence of non-AD dementia as $P(Z^{(1)}(t)=2\mid Z^{(1)}(0)=1)+P(Z^{(1)}(t)=4\mid Z^{(1)}(0)=1)$.
For type II disease, the cumulative incidence of AD is $P(Z^{(2)}(t)=3\mid Z^{(2)}(0)=1)+P(Z^{(2)}(t)=6\mid Z^{(2)}(0)=1)$.

\section{Details of the Bayesian analyses of the Nun Study}

In the scenario with a non-informative prior, we utilized a Uniform (0,1) distribution for the prior of the initial probabilities $\psi_m$ and $\boldsymbol{\pi}^{(m)}$, with the constraint $\sum_{j=1}^{J^m}\pi_j^{(m)}=1$. For the logarithm of the transition intensities, we applied a Normal distribution with mean 0 and variance 1000.
To incorporate prior information from the ADAMS study, we assigned the following prior distributions: $\textrm{logit }\psi_2\sim N(0.99, 2)$, $\textrm{logit }\pi^{(1)}_2\sim N(-3.69, 0.5)$ and $\textrm{logit }\pi^{(2)}_3\sim N(-3.74, 0.5)$. Other priors and constraints remained the same as the non-informative scenario.
Table~\ref{tb:7} gives the Bayesian estimates using both a non-informative prior and the informative prior based on the ADAMS study as described above. 

\begin{table}[h!]
	\caption{Bayesian analyses for the Nun Study using non-informative prior and prior information from the ADAMS. \\
		Est: mean of the posterior distribution; $\widehat{R}$: potential scale reduction factor;
		$N_{\textrm{eff}}$: effective number of iterations.
  ADAMS: Aging, Demographics, and Memory Study \citep{plassman2007prevalence}.}
	
	\label{tb:7}
	
	\centering
	\begin{tabular}{c rrrr rrrr}
		\hline \\[-1em]
		&  \multicolumn{4}{c}{Non-informative Prior} & \multicolumn{4}{c}{ADAMS Prior}\\ 	\\[-1em]
		& Est & 95\% CI & $\widehat{R}$ & $N_{eff}$ & Est & 95\% CI & $\widehat{R}$ & $N_{eff}$ \\ 
		\hline 
		\\[-1em]
		
		$\lambda^{(1)}_{12}$ & 0.148 & (0.102, 0.202) & 1.001 & 2000 & 0.044 & (0.030, 0.059) & 1.003 & 610 \\ \\[-1em]
		$\lambda^{(1)}_{13}$ & 0.000 & (0.000, 0.001) & 1.001 & 2000 & 0.099 & (0.076, 0.124) & 1.004 & 2000 \\ \\[-1em]
		$\lambda^{(1)}_{24}$& 0.404 & (0.314, 0.500)& 1.003 & 1900 & 0.374 & (0.284,  0.474) & 1.002 & 1300 \\ \\[-1em]
		
		$\lambda^{(2)}_{13}$ & 0.101 & (0.088, 0.116) & 1.002 & 860 & 0.202 & (0.165, 0.254) & 1.006 & 290 \\ \\[-1em]
		$\lambda^{(2)}_{14}$ & 0.066 & (0.051, 0.083) & 1.001 & 2000 & 0.000 & (0.000, 0.003) & 1.001 & 2000 \\ \\[-1em]
		$\lambda^{(2)}_{23}$ & 0.133 & (0.107, 0.167)& 1.002 & 1000 & 0.116 & (0.093, 0.146) & 1.015 & 160 \\ \\[-1em]
		$\lambda^{(2)}_{25}$ & 0.095 & (0.072, 0.125) & 1.001 & 1900 & 0.084 & (0.063, 0.109)& 1.003 & 600 \\ \\[-1em]
		$\lambda^{(2)}_{36}$ & 0.252 & (0.222, 0.282) & 1.001 & 2000 & 0.256 & (0.227, 0.288)& 1.001 & 2000 \\ \\[-1em]
		
		$\psi_2$ & 0.871 & (0.833, 0.905)& 1.002 & 2000 & 0.682 & (0.637, 0.723) & 1.002 & 2000 \\ \\[-1em]
		$\pi^{(1)}_{1}$ & 0.782 & (0.478, 0.990) & 1.018 & 200 & 0.973 & (0.926, 0.994)& 1.001 & 2000 \\ \\[-1em]
		$\pi^{(2)}_{1}$ & 0.969 & (0.901, 0.998) & 1.012 & 300 & 0.929 & (0.816, 0.982) & 1.001 & 2000 \\ \\[-1em]
		$\pi^{(2)}_{2}$ & 0.015 & (0.000, 0.070) & 1.005 & 370 & 0.046 & (0.001, 0.160)& 1.002 & 2000 \\ \\[-1em]
		\hline
	\end{tabular}
\end{table}

For numerical computations, we employed the Gibbs Sampler with 2000 iterations, implemented using the R function``jags()" in the R2jags package \citep{R2jags}.  
Convergence of the Gibbs Sampler is evaluated by the effective number of iterations $N_{\textrm{eff}}$ and the potential scale reduction factor $\widehat{R}$ \citep{gentleman1994}. 
Generally, successful convergence is indicated by $N_{\textrm{eff}}\geq30$ and $\widehat{R}\approx1$.

\end{document}